# Measuring lumbar back motion during functional activities using a portable strain gauge sensor-based system: a comparative evaluation and reliability study

Magdalena SUTER[1], Patric EICHELBERGER[1], Jana FRANGI[1], Edwige SIMONET[1], Heiner BAUR[1], Stefan SCHMID[1,*]

[1]Bern University of Applied Sciences, Department of Health Professions, Division of Physiotherapy, Spinal Movement Biomechanics Group, Bern, Switzerland

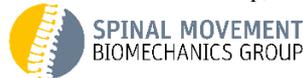

**Corresponding author:**
[*]Stefan Schmid, PT, PhD, Bern University of Applied Sciences, Department of Health Professions, Murtenstrasse 10, 3008 Bern, Switzerland, +41 79 936 74 79, stefanschmid79@gmail.com

**ABSTRACT**

Quantifying lumbar back motion during functional activities in real-life environments may contribute to a better understanding of common pathologies such as spinal disorders. The current study therefore aimed at the comparative evaluation of the Epionics SPINE system, a portable device for measuring sagittal lumbar back motion during functional activities. Twenty healthy participants were therefore evaluated with the Epionics SPINE and a Vicon motion capture system in two identical separate research visits. They performed the following activities: standing, sitting, chair rising, box lifting, walking, running and a counter movement jump (CMJ). Lumbar lordosis angles were extracted as continuous values as well as average and range of motion (ROM) parameters. Agreement between the systems was evaluated using Bland-Altman analyses, whereas within- and between-session reliability were assessed using intraclass correlation coefficients (ICC) and minimal detectable changes (MDC). The analysis showed excellent agreement between the systems for chair rising, box lifting and CMJ with a systematic underestimation of lumbar lordosis angles during walking and running. Reliability was moderate to high for all continuous and discrete parameters (ICC≥0.62), except for ROM during running (ICC=0.29). MDC values were generally below 15°, except for CMJ (peak values up to 20° within and 25° between the sessions). The Epionics SPINE system performed similarly to a Vicon motion capture system for measuring lumbar lordosis angles during functional activities and showed high consistency within and between measurement sessions. These findings can serve researchers and clinicians as a bench mark for future investigations using the system in populations with spinal pathologies.

# 1. INTRODUCTION

The evaluation of lumbar back motion during daily activities is of high importance to better understand the pathomechanics of common pathologies such as non-specific chronic low back pain (NSCLBP) (Christe et al., 2017; Christe et al., 2016; Gombatto et al., 2015; Hemming et al., 2018; Hernandez et al., 2017; Papi et al., 2019). This is commonly achieved using complex laboratory-based measurement approaches such optical motion capturing, which is costly, time consuming and thus not available in many clinical settings (Alqhtani et al., 2015; Pfister et al., 2014). As an alternative, portable and wearable sensor systems have been introduced, allowing a cost- and time-efficient assessment of body movement in real-life environments (Papi et al., 2017). Thereby, while inertial measurement unit (IMU)-based systems are known to be affected by drift errors when collecting data over longer time periods (Bergamini et al., 2014), strain gauge sensor-based systems such as the Epionics SPINE system might be more appropriate for measuring lumbar back motion during daily activities. With the capacity of collecting motion data for up to 24 hours, this system was previously used to investigate lumbar back alignment and motion over the course of a full day (Dreischarf et al., 2016; Rohlmann et al., 2014). Moreover, its easy-to-apply design facilitates data collections in large cohorts (Consmüller et al., 2012b; Consmüller et al., 2014; Dreischarf et al., 2014; Pries et al., 2015; Schmidt et al., 2018a; Schmidt et al., 2018b). However, despite two studies involving comparative evaluations and consistency analyses of the Epionics SPINE system during standing as well as isolated flexions and extensions (Consmüller et al., 2012a; Taylor et al., 2010), it remains unclear how accurate and consistent the system measures lumbar curvature during repeated functional activities. In addition, these studies evaluated static postural angles and ROM parameters and did not include evaluations of the continuous motion data, which is important when aiming at a particular phase of a movement, e.g. the loading response or push-off phases during gait.

For these reasons, the aim of this study was to compare continuous and discrete (parameterized as average and ROM) lumbar lordosis angles during functional activities measured with the Epionics SPINE system to those measured with a Vicon motion capture system and to evaluate the reliability of continuous and discrete lumbar lordosis angles measured with the Epionics SPINE system within a measurement session and between two measurement sessions separated by about a week.

# 2. METHODS

## 2.1. Participants

Twenty healthy adults (11 females/9 males; height: 173±10 (157-192) cm; mass: 69±13 (45.5-91.7) kg; age: 31±9 (20-53) years; body mass index (BMI): 22.6±2.6 (18.5-27.5) kg/m$^2$) participated in the current study. They were recruited by flyer and inquiries from the community surrounding the authors' institution. Exclusion criteria were obesity (BMI of greater than 30kg/m$^2$), anamnestically known pregnancy, psychological or neuromuscular disorders, spinal pathologies, nerve root pain or any other injuries or surgeries to the locomotor system which restricted normal movement. The protocol was approved by the local ethics committee (Kantonale Ethikkommission des Kantons Bern, protocol no.: 2017-00820) and written informed consent was obtained from each participant. The sample size of twenty was chosen based on the range of sample sizes described throughout the scientific literature on movement analysis reliability (Benedetti et al., 1998; Kaufman et al., 2016; Wilken et al., 2012).

## 2.2. Measurement systems

### 2.2.1. Epionics SPINE system

The Epionics SPINE system (Epionics Medical GmbH, Potsdam, Germany) is composed of two flexible sensor stripes, each of which including 12 same-sized strain gauge sensors that register bending angles on the back surface at a sampling rate of 50Hz. For each participant, the sensor stripes were inserted in two special concave tapes that were attached to the skin 3 cm lateral and parallel on both sides of the spinous processes, starting from the height of the posterior superior iliac spines (Figure 1). Two cables connected the sensor stripes to a portable storage unit (size: 12.5 cm x 5.5 cm, mass: 80 g). To ensure smooth gliding of the sensor stripes within the concave tapes, female participants were provided an individually adjusted sports bra.

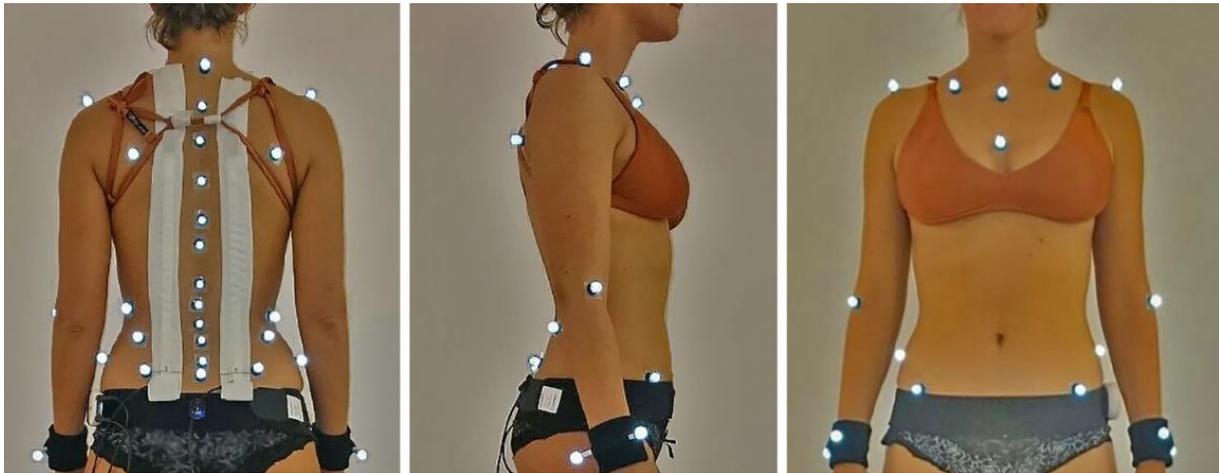

**Figure 1:** Attachment of the retro-reflective markers (only shown for the trunk) as well as the paravertebral strain gauge sensor stripes and the corresponding storage unit of the Epionics SPINE system.

### 2.2.2. Optical motion capture system

Comparative data were recorded using a 10-camera optical motion capture system (Vicon, Oxford, UK; sampling frequency: 200 Hz) and a previously described marker configuration (Schmid et al., 2017). In brief, 56 retro-reflective markers were placed according to the Plug-in Gait full body (Schweizer et al., 2014) and the IfB trunk marker sets (List et al., 2013), which included markers on the spinous processes of the vertebrae C7, T3, T5, T7, T9, T11, L1-L5 and the sacrum (Figure 1).

## 2.3. Experimental procedures

Data collection took place in a movement analysis laboratory. Participants were invited for two identical research visits, which were separated by 7-10 days from each other. At each visit, an experienced physiotherapist performed a standard clinical examination to ensure normal joint mobility and muscle strength and equipped the participant with the markers as well as the sensor stripes.
Subsequently, they were asked to stand and sit for 10 seconds each in an upright position with the arms hanging relaxed at the sides and to perform the following functional activities: 1) Standing up from a chair with free arms; 2) Lifting up a light-weighted box (5 kg) in front of the body by squatting down; 3) Walking and running on a 10-meter level ground; 4)

Performing a vertical counter movement jump (CMJ) from a standing position using the "simultaneous arm swing" technique (Gutierrez-Davila et al., 2014). Chair height was adjusted for each participant so that the hip and knee joints were flexed about 90 degrees (Schmid et al., 2013). All measurements were conducted barefoot and the functional activities were performed at a self-selected normal speed and repeated until at least four valid trials were collected. Every activity was explained and demonstrated by an investigator and the participants were given as many practice trials as necessary. To synchronize the two systems, participants were required to perform a rapid lumbar flexion in a standing position at the beginning of the measurement of each functional activity (reference movement).

**2.4. Data reduction and parameters of interest**

Pre-processing of the motion capture data (i.e. reconstruction, labeling and filtering of the marker trajectories as well as setting temporal events to identify the relevant data sections) was implemented with the software Nexus (version 2.6, Vicon UK, Oxford, UK). Post-processing of the motion capture data as well as processing of the Epionics SPINE data was carried out using a custom-built MATLAB-routine (R2018b, MathWorks Inc., Natrick, MA, USA).

In a first step, raw data from both systems were extracted and used to calculate sagittal lumbar curvature angles. For the Epionics SPINE system, curvature angles were calculated by averaging the sums of the bending angles acquired by each strain gauge sensor located below the spinous process of T12 on the left and right sides of the spine and up-sampled to 200 Hz using linear interpolation to match the sampling frequency of the motion capture data. Motion capture-based curvature angles were established using a combination of a $2^{nd}$ order polynomial and a circle fit function that was applied to the trajectories of the markers placed on the spinous processes of L1-5 and the sacrum. Details on curvature angle calculation, marker placement accuracy and soft tissue artifacts can be found elsewhere (Schmid et al., 2015; Zemp et al., 2014).

In order to extract the relevant movement cycles, data from the two measurement systems were synchronized using the peak flexion lumbar curvature angle from the reference movement and cut according to the temporal events that were set based on the motion capture data (Figure 2). The following events were thereby accepted as set during data pre-processing: beginning of box lifting movement (instant when the box was lifted off the ground, identified using force plate where the box was placed on), beginning and end of the gait and running cycles (two sequential heel strikes on the left side) as well as end of CMJ (touch down after flight phase, identified using force plate under the participants' feet). All other events were defined using an event detection function. The beginning of the chair rising movement was defined by the instant when the forward motion of the sternum marker exceeded 20 mm from the initial position. The end of the chair rising as well as box lifting movements were defined as the instant when the sacrum marker moved forward within 20 mm of the final position. The beginning of the CMJ was defined by the instant when the backward motion of the sacrum marker exceeded 20 mm from the initial position.

Lumbar curvature angles from both systems were then low-pass filtered at a cutoff frequency of 6 Hz (Butterworth, fourth order, zero-phase) and time normalized to movement cycles consisting of 101 data points. In addition, continuous data were parameterized into average and ROM values. Prior to any further analyses, all data were checked for plausibility and erroneous trials were excluded.

**2.5. Statistical analyses**

Statistical analyses were conducted using MATLAB (R2018b, MathWorks Inc., Natrick, MA, USA) and the MATLAB software package for one-dimensional Statistical Parametric

Mapping (SPM: spm1d-package, www.spm1d.org) (Pataky, 2012). Normality was confirmed by the Kolmogorov-Smirnov test for the discrete parameters and the SPM function *spm1d.stats.normality.ttest* (Chi-square test) for the continuous data using an alpha-level of 0.05 for both tests. To compare the lumbar curvature angles acquired with the Epionics SPINE system to those acquired with the Vicon motion capture system, Bland-Altman analyses with mean differences (Epionics minus Vicon) and 95% limits of agreement (LoA) were performed (Bland and Altman, 2003). Additionally, one sample T-tests with an alpha-level of 0.05 were used to explore the deviation of the mean differences from 0. Within- and between-session reliability was assessed according to the three-layered approach suggested by Weir (2005). This included the evaluation of 1) systematic errors by comparing the average of the individual differences to 0 using one samples T-tests with an alpha-level of 0.05, 2) relative reliability using intraclass correlation coefficients (consistency formula, ICC(C,1)), and 3) absolute reliability using minimal detectable changes (MDC). MDC was thereby calculated as 1.96*SDd (standard deviation of the differences), which represents the smallest degree of change that exceeds measurement error and can be used to distinguish true changes from changes caused by errors. One sample T-tests for the continuous data were implemented using the SPM function *spm1d.stats.ttest*, which calculates the t-statistics for the respective group comparison and uses Random Field Theory (RFT) (Adler and Taylor, 2007) to identify so called "supra-threshold clusters", i.e. clusters where the t-value crosses a critical threshold that corresponds to a pre-defined alpha-level. To enable the straightforward interpretation of the identified group differences over time, the RFT procedures ultimately provide a single p-value for each supra-threshold cluster (Pataky, 2012).

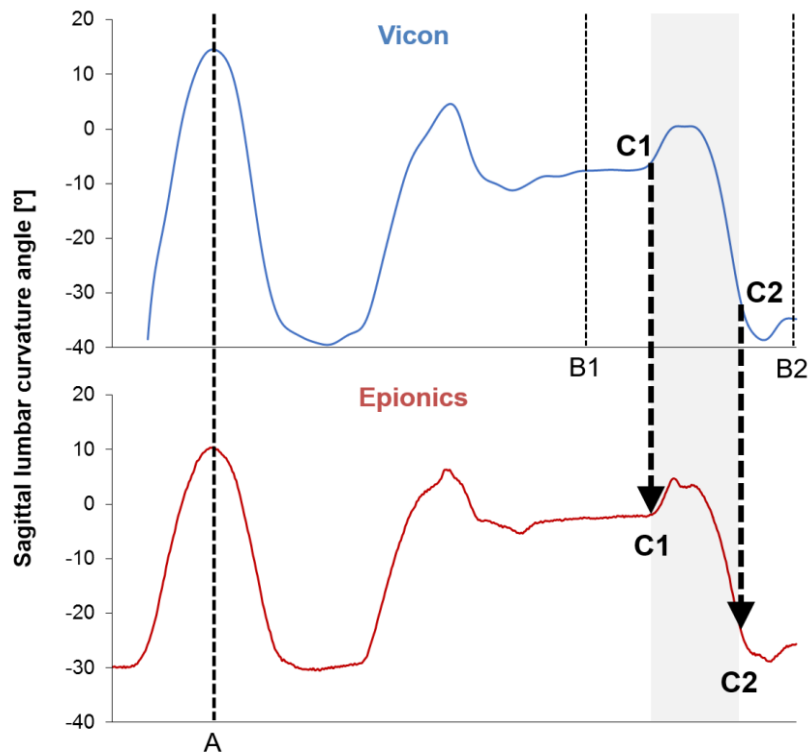

**Figure 2:** Synchronization of the Epionics SPINE and Vicon motion capture systems on the basis of the peak flexion lumbar curvature angle from the reference movement (A) using the example of chair rising. The points B1 and B2 illustrate the temporal events set manually during pre-processing of the motion capture data. Using a custom event detection algorithm, they were used to identify the beginning (C1) and end (C2) of the respective movement cycle (gray shaded area).

## 3. RESULTS

### 3.1. Discrete parameters

Although the deviations of the lumbar lordosis angle mean differences from 0 appeared to be beyond chance variation for the average values of standing, walking and running (Epionics measurements indicating less average lumbar lordosis by 7.0°, 11.0° and 6.5°, respectively; p≤0.031) as well as the ROM value of running (Epionics measurements indicating less lumbar lordosis ROM by 4.0°; p=0.016), Bland-Altman analyses indicated generally high agreement between the two systems (Table 1). The lower LoAs ranged from -4.7° for walking (average value) to -26° for CMJ (ROM value), whereas upper LoAs were between 6.6° for running (ROM value) and 28° for standing and walking (average values).

As for reliability, apart from the ROM value of running within the measurement session (p=0.034), no systematic errors could be identified. ICCs of greater or equal to 0.78 indicated high consistency within the measurement session, with MDCs ranging from 2.5° to 8.1°, except from the ROM value of CMJ (14.0°). Between the measurement sessions, moderate to low consistency was found for the ROM values of chair rising (ICC=0.62), CMJ (ICC=0.60), box lifting (ICC=0.50) and running (ICC=0.29), whereas all other parameters showed high consistency (ICC≥0.77) with MDCs ranging from 1.9° for walking (ROM value) to 12.0° for sitting (average value).

**Table 1:** Results for the comparative evaluation (Epionics SPINE system vs. Vicon motion analysis system) as well as within-session and between-session reliability of the lumbar lordosis angle (reported in degrees).

| Activity | Parameter | Comparative evaluation | | | | | Within-session reliability | | | Between-session reliability | | |
|---|---|---|---|---|---|---|---|---|---|---|---|---|
| | | Mean lordosis angle (SD)[1] | | Bland-Altman analysis | | | T-test | T-test | ICC | MDC | T-test | ICC | MDC |
| | | Vicon | Epionics | Mean diff.[2] | Upper LoA | Lower LoA | p-value | p-value | | | p-value | | |
| Standing | Average | -45.5 (15.1) | -38.4 (10.5) | 7.0 | 28.0 | -14.0 | 0.031* | N/A | N/A | N/A | 0.510 | 0.94 | 6.6 |
| Sitting | Average | -16.7 (11.4) | -16.1 (9.9) | 0.7 | 10.0 | -9.1 | 0.672 | N/A | N/A | N/A | 0.381 | 0.82 | 12.0 |
| Walking | Average | -45.7 (13.1) | -34.3 (9.8) | 11.0 | 28.0 | -4.7 | >0.001* | 0.414 | 0.99 | 3.6 | 0.610 | 0.86 | 11.0 |
| | ROM | -8.8 (5.6) | -6.1 (2.3) | -2.6 | 7.1 | -12.0 | 0.093 | 0.952 | 0.78 | 3.3 | 0.749 | 0.92 | 1.9 |
| Running | Average | -38.3 (13.1) | -31.8 (8.9) | 6.5 | 22.0 | -9.0 | 0.008* | 0.579 | 0.99 | 2.5 | 0.457 | 0.87 | 11.0 |
| | ROM | 12.3 (6.9) | 8.3 (2.8) | -4.0 | 6.6 | -15.0 | 0.016* | 0.034* | 0.94 | 3.1 | 0.952 | 0.29 | 10.0 |
| Chair rising | Average | -19.4 (12.6) | -19.0 (9.3) | 0.5 | 12.0 | -11.0 | 0.743 | 0.624 | 0.96 | 5.6 | 0.367 | 0.88 | 9.3 |
| | ROM | 30.5 (9.8) | 26.6 (6.5) | -3.9 | 13.0 | -21.0 | 0.061 | 0.130 | 0.86 | 7.1 | 0.978 | 0.62 | 8.8 |
| Box lifting | Average | -14.9 (9.6) | -16.0 (7.8) | -1.1 | 10.0 | -13.0 | 0.421 | 0.228 | 0.93 | 6.1 | 0.257 | 0.77 | 11.0 |
| | ROM | 37.2 (11.4) | 36.4 (6.8) | -0.83 | 20.0 | -22.0 | 0.737 | 0.300 | 0.84 | 8.1 | 0.900 | 0.50 | 12.0 |
| CMJ | Average | -26.4 (15.2) | -24.5 (9.9) | 1.9 | 18.0 | -14.0 | 0.352 | 0.898 | 0.97 | 5.0 | 0.475 | 0.90 | 9.1 |
| | ROM | 41.5 (11.7) | 38.2 (10.6) | -3.3 | 19.0 | -26.0 | 0.236 | 0.825 | 0.81 | 14.0 | 0.484 | 0.61 | 18.0 |

[1] Negative average values represent a lordotic posture of the lumbar back region.
[2] Differences between the mean lordosis angles acquired by the Epionics SPINE and Vicon motion capture systems (Epionics-Vicon).
Abbreviations: CMJ=Countermovement jump; ROM=Range of motion; LoA=Limits of agreement; ICC=Intraclass correlation coefficient; MDC=Minimal detectable change.

### 3.2. Continuous data

Overall, agreement between the two systems appeared equally high for the continuous data (Figure 3). Although the Epionics SPINE system tended to systematically underestimate the lumbar curvature angles by about 10° for walking (supra-threshold cluster identified at 0-100% of gait cycle; p<0.001) and 5° for running (supra-threshold clusters identified at 28-49% and 79-94% of running cycle; p=0.016 and p=0.029, respectively), all other mean differences seemed to be due to chance only (Figure 3, right column). LoA indicated slightly lower agreement during the stance phases of running, the first 20% of chair rising as well between 0-10% and approximately 40-90% of the CMJ (Figure 3, middle column).

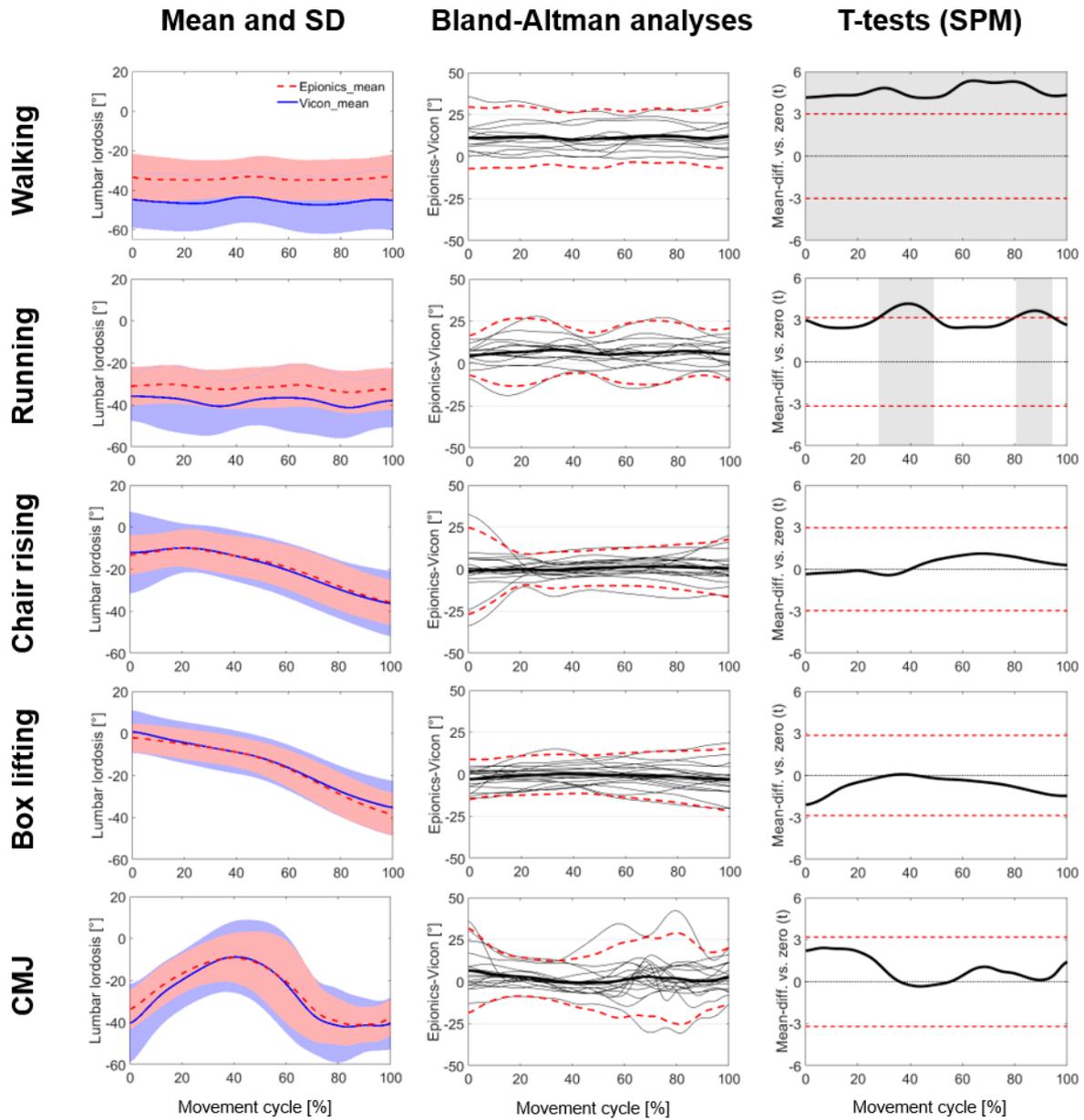

**Figure 3:** Comparison between continuous lumbar lordosis angles acquired with the Epionics SPINE system (red) and those acquired with the Vicon motion capture system (blue) for the activities walking, running, chair rising, box lifting and counter movement jump (CMJ). The left column shows mean and standard deviation (SD, shaded areas) of the lumbar lordosis angles measured with the respective system. The middle column illustrates the results of the Bland-Altman analyses with mean differences (black solid lines) and 95% limits of agreement (red dotted lines). The right column illustrates the results of the comparisons between the overall mean differences and 0 using one-dimensional Statistical Parametric Mapping (SPM), with the red dotted lines indicating the critical thresholds for statistical significance at the p=0.05 level and the gray shaded areas the supra-threshold clusters, where the t-value crosses the critical threshold indicating statistically significant deviations of the mean differences from 0.

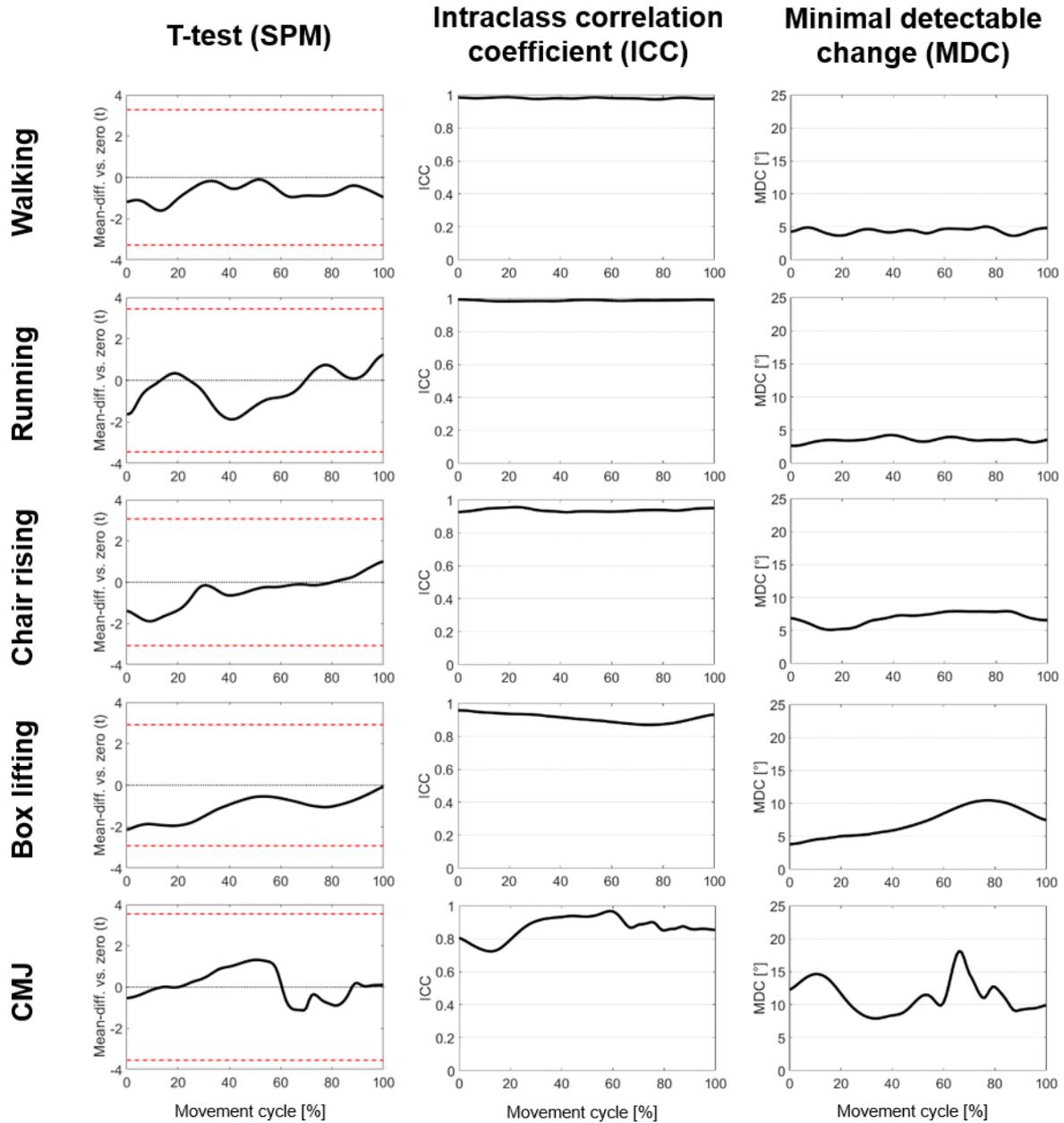

**Figure 4:** Within-session reliability for continuous lumbar lordosis angles acquired with the Epionics SPINE system for the activities walking, running, chair rising, box lifting and counter movement jump (CMJ). The left column illustrates the results of the evaluation for systematic errors using independent samples T-tests (implemented by means of one-dimensional Statistical Parametric Mapping, SPM), with the red dotted lines indicating the thresholds for statistical significance at the p=0.05 level. The middle and right columns show the intraclass correlation coefficients (ICC, consistency formula) for relative reliability and minimal detectable changes (MDC) for absolute reliability, respectively.

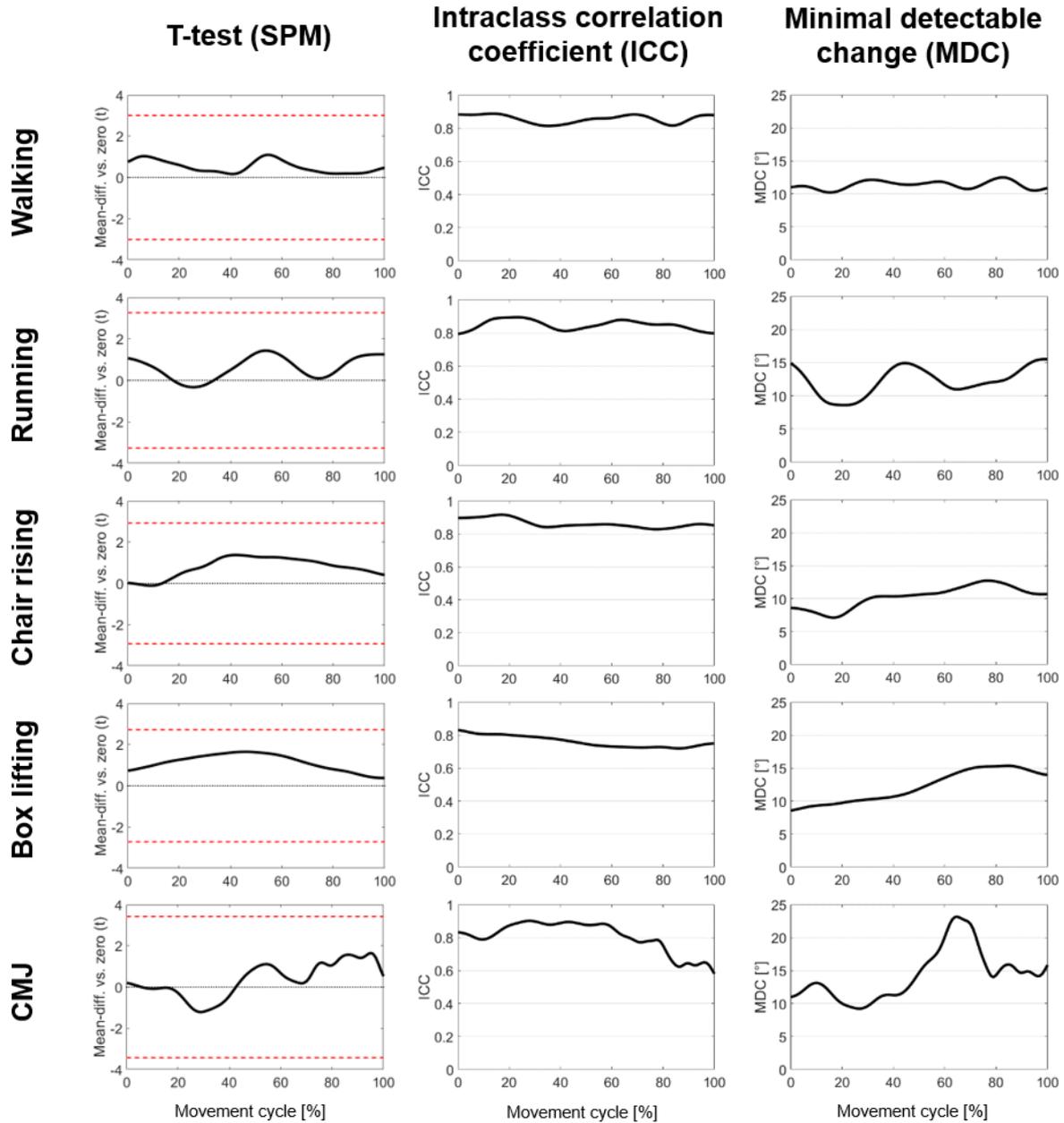

**Figure 5:** Between-session reliability for continuous lumbar lordosis angles acquired with the Epionics SPINE system for the activities walking, running, chair rising, box lifting and counter movement jump (CMJ). The left column illustrates the results of the evaluation for systematic errors using independent samples T-tests (implemented by means of one-dimensional Statistical Parametric Mapping, SPM), with the red dotted lines indicating the thresholds for statistical significance at the p=0.05 level. The middle and right columns show the intraclass correlation coefficients (ICC, consistency formula) for relative reliability and minimal detectable changes (MDC) for absolute reliability, respectively.

Considering reliability, no systematic errors were found (i.e. no supra-threshold clusters were identified) within and between the measurement sessions (Figures 4 and 5, left column). ICCs of greater than 0.8 indicated very high within-session consistency, except for the first 20% of the CMJ cycle, where ICC reached a minimum of 0.72 (Figure 4, middle column). MDCs ranged from 2.5° to 10.5°, with peak values reaching about 18.0° around 60-70% of the CMJ cycle (Figure 4, right column).

Consistency was similarly high between the measurement sessions, except for box lifting around 40-100% and CMJ around 65-100% of the movement cycle (ICC reaching minimum values of 0.72 and 0.58, respectively) (Figure 5, middle column). MDCs were generally higher than those within the measurement session, with peak values reaching about 23.0° for CMJ around 70% of the movement cycle (Figure 5, right column).

## 4. DISCUSSION

The current study aimed at the comparison of discrete (average and ROM) and continuous sagittal lumbar curvature angles measured with the Epionics SPINE system during functional activities to those measured with a Vicon motion capture system. Moreover, consistency of the Epionics SPINE system-based angles were assessed within a single and between two separate measurement sessions. The analysis revealed high agreement between the two systems, with a tendency for systematic underestimation of the Epionics SPINE system-based lumbar curvature angles during standing, walking and running. Consistency within the measurement session was found to be high across all activities and parameters. Between the measurement sessions, the results showed high consistency for all average and most continuous parameters, but only low to moderate consistency for the majority of ROM parameters as well as continuous lumbar curvature angles during the flight phase of the CMJ.

This is the first study involving a comprehensive evaluation of the Epionics SPINE system for the dynamic assessment of lumbar lordosis angles during functional activities across separate measurement sessions. The findings are in line with previous studies evaluating accuracy and repeatability of sagittal lumbar curvature angles with the current and earlier versions of the Epionics SPINE system during standing as well as isolated flexion and extension (Consmüller et al., 2012a; Taylor et al., 2010).

The fact that the Epionics SPINE system tended to underestimate lumbar lordosis angles during standing, walking and running, but not during sitting, could be explained by greater lordosis angles during upright standing activities. It was previously shown that lumbar lordosis angles of more than 40° were less accurately estimated by motion capture-based surface measurements than radiographic data, which was suggested to be due to soft tissue accumulation with increasing lordosis (Schmid et al., 2015).

The limited agreement for lumbar curvature angle ROM during running might be explained by a delayed gliding of the sensor stripes within the concave tapes, especially during rapid directional changes of lumbar motion. In addition, while the evaluation with the Epionics SPINE system solely focused on bending angles in the sagittal plane, locomotion activities also include lateral bending and axial rotational motion, which may have had an influence on lumbar ROM.

The slightly lower reliability for the CMJ, especially between the two separate measurement sessions, might be associated with the chosen level of standardization. Previous studies discussed the standardization of jumping, for example in terms of reach height (Ferreira et al., 2010; Meylan et al., 2010). Our current approach was to give clear instructions on how the movement should be performed but at the same time allow for variation and individual strategies, which is of particular importance when investigating movement behavior in pathologies such as NSCLBP. Too much standardization would thereby most likely cause

"wash out" effects and impede the identification of mechanical factors that might contribute to the pathology.

The rather wide LoAs and high MDC values during the push-off phase of the CMJ might be related to the system's limited sampling frequency of 50Hz during high-velocity curvature angle changes. In addition, curvature changes at such high velocities are heavily dependent on the definition of beginning and end of the movement. If the movement initiation is slightly different from trial to trial, this would likely result in time shifts that would be particularly visible during these phases. Together with the possibly restricted gliding of the sensor stripes within the concave tapes, this indicates that the system might be limited for quantifying lumbar back motion during high-speed movements such as running and jumping.

However, the goal of this study was not to classify MDC values as high or low, but rather to provide a bench mark, which can serve future studies using the Epionics SPINE system for the evaluation of functional activities to ensure that a difference captured between two measures can be considered "real" and is not just due to measurement noise. Such knowledge is necessary when investigating treatment effects, especially in situations where probability calculations indicate statistical significance for an observed difference. In addition, MDC values should always be considered when determining whether an observed difference is of clinical importance. The smallest amount of change in a parameter that is considered relevant or important to patients or clinicians must thereby be larger than the MDC to be useful (Stipancic et al., 2018). For example, even though clinicians might consider a change in lumbar lordosis of 10° as clinically relevant, it should not be further interpreted when MDC is 15° for a particular measurement approach.

A strength of this study was the comprehensive evaluation of agreement and reliability for continuous angle data, which provides more specific information for future studies targeting the analysis of movement characteristics during selected phases of a movement cycle that could not be appropriately quantified using discrete parameters such as average or ROM.

The fact that motion capture system was not always able to fully identify and distinguish the lumbar markers (especially in smaller participants) was considered a limitation of this study. Most missing data was thereby found for the walking and running activities, which can again be associated with the increased lordotic posture in upright standing activities. Future studies using similar motion capture approaches might address this issue by using more cameras, cameras with a higher resolution and/or less lumbar markers. Another limiting factor of this study was that we did not include individuals with a BMI of greater than 30kg/m$^2$. We have chosen this particular cutoff value to exclude individuals with obesity, which we considered an unhealthy condition and hence an exclusion criterion for our study population. However, BMI alone might not be adequate to decide whether the Epionics SPINE system is applicable or not. More importantly thereby might be to consider information on body composition, e.g. whether a high BMI is dependent on a large amount of visceral or evenly distributed subcutaneous adipose tissue. Finally, the fact that the Vicon markers were not placed on the same anatomical landmarks as the sensor stripes of the Epionics SPINE system might have slightly limited the comparative evaluation. However, placing the markers on top of the concave tapes in which the sensor stripes were inserted would most likely have resulted in a larger error than placing the markers 3 cm apart from the sensor stripes on the spinous processes.

In conclusion, the Epionics SPINE system performed similarly to a Vicon motion capture system for measuring sagittal lumbar back angles during functional activities and showed high consistency within and between measurement sessions. These findings can serve researchers and clinicians as a bench mark for future investigations using the Epionics SPINE system in populations with spinal pathologies. Careful use of the system is thereby recommended when investigating movements that involve fast directional changes of lumbar lordosis angles such as running or jumping.

## 5. CONFLICT OF INTEREST STATEMENT

The authors declare no conflicts of interest.

## 6. ACKNOWLEDGMENTS

The authors thank Andre Kwiatek from Epionics Medical GmbH in Potsdam, Germany for providing us with the Epionics SPINE system at no charge as well as Dr. Michael L. Meier and Fabienne Riner for their assistance in the pre-processing of the motion capture data. This study was partially funded by the Swiss Physiotherapy Association (physioswiss).